\documentclass[12pt]{article}
\usepackage{a4wide}
\usepackage{epsfig}
\usepackage{color}

\newcommand{\slk}{/\kern-6pt k}
\newcommand{\sll}{/\kern-4pt l}
\newcommand{\slp}{p\kern-5pt/}
\newcommand{\slq}{q\kern-5.5pt/}
\newcommand{\sls}{s\kern-5.5pt/}
\newcommand{\tr}{\mathop{\rm tr}\nolimits}
\newcommand{\Tr}{\mathop{\rm Tr}\nolimits}
\newcommand{\bea}{\begin{eqnarray}}
\newcommand{\ena}{\end{eqnarray}}
\newcommand{\nn}{\nonumber\\}
\newcommand{\be}{\begin{equation}}
\newcommand{\en}{\end{equation}}
\newcommand{\oone}{\hbox{$1\kern-2.5pt\hbox{\rm l}$}}
\newcommand{\ssigma}{\hbox{$\kern2.5pt\vrule height4pt\kern-2.5pt\sigma$}}

\newcommand{\GeV}{{\rm\,GeV}}
\newcommand{\tfrac}[2]{{\textstyle\frac{#1}{#2}}}

\newcommand{\Li}{{\rm Li}}
\newcommand{\imag}{\mathop{\rm Im}\nolimits}
\newcommand{\real}{\mathop{\rm Re}\nolimits}

\newcommand{\slell}{/\kern-5pt\ell}

\begin{document}

\thispagestyle{empty} 
\begin{flushright}
MITP/18-002
\end{flushright}
\vspace{0.5cm}

\begin{center}
{\Large\bf $T$-odd correlations in polarized top quark decays in the
sequential decay $t(\uparrow) \to X_b+W^+(\to \ell^+ + \nu_\ell)$ and in the
quasi-three-body decay $t(\uparrow) \to X_b+ \ell^+ + \nu_\ell$}\\[1.3cm]
{\large M.~Fischer$^1$,  S.~Groote$^2$ and J.G.~K\"orner$^3$ }\\[1cm]
$^1$ Keltenstr.~18, 64625 Bensheim, Germany\\[7pt]
$^2$ Loodus- ja t\"appisteaduste valdkond, F\"u\"usika Instituut,\\[.2cm]
  Tartu \"Ulikool, W.~Ostwaldi~1, 50411 Tartu, Estonia\\[7pt]
$^3$ PRISMA Cluster of Excellence, Institut f\"ur Physik,  \\[.2cm]
  Johannes-Gutenberg-Universit\"at, D-55099 Mainz, Germany\\[7pt]
 \end{center}

\vspace{1cm}
\begin{abstract}\noindent
We identify the $T$-odd structure functions that appear in the description
of polarized top quark decays in the sequential decay
$t(\uparrow) \to X_b+W^+(\to \ell^+ + \nu_\ell)$ (two structure functions) and
the quasi-three-body decay $t(\uparrow) \to X_b+ \ell^+ + \nu_\ell$
(one structure function). A convenient measure of the magnitude of the $T$-odd
structure functions is the contribution of the imaginary part $\imag g_R$ of
the right-chiral tensor coupling $g_R$ to the $T$-odd structure functions
which we work out. Contrary to the case of QCD the NLO electroweak corrections
to polarized top quark decays admit of absorptive one-loop vertex
contributions. We analytically calculate the imaginary parts of the relevant
four electroweak one-loop triangle vertex diagrams and determine their
contributions to the $T$-odd helicity structure functions that appear in the
description of polarized top quark decays.
\end{abstract}

\section{Introduction}
Large numbers of single top quarks have been and are being currently produced
at the LHC~\cite{Aaboud:2016ymp,Aaboud:2017pdi,Sirunyan:2016cdg,CMS:2016xnv}.
The present situation concerning both ATLAS and CMS results on single top
production is nicely summarized in a review article by
N.~Faltermann~\cite{Faltermann:2017vry}. The dominant production mechanism is
the so-called $t$-channel production process. The production proceeds via
parity-violating weak interactions -- a necessary condition for the top quark
to be polarized. In fact, theoretical calculations predict an average
polarization close to $90\%$~\cite{Mahlon:1999gz,Schwienhorst:2010je} where
the polarization is primarily along the direction of the spectator quark.
The polarization of singly produced top quarks has been measured by the
CMS Collaboration~\cite{Khachatryan:2015dzz} ($P_t=0.58\pm0.22$), by the ATLAS
Collaboration~\cite{Aaboud:2017aqp} ($P_t=0.97\pm0.12$) and, most recently,
again by the ATLAS collaboration who quote a polarization value of
$|\vec P|> 0.72$ at a confidence level of $95\%$~\cite{Aaboud:2017yqf}.

There are two ways in which polarized top quark decays can be analyzed. In the
first approach one first considers the quasi-two-body decay
$t(\uparrow) \to X_b + W^+$ which is analyzed in the top quark rest frame. The
subsequent decay $W^+ \to \ell^+ + \nu_\ell$ is analyzed in the $W^+$ rest
frame. One first calculates the spin density matrix elements of the produced
gauge boson $W^+$ in the production process $t \to X_b +W^+$ and then analyzes
the spin density matrix with the help of the decay $W^+ \to \ell^+ +\nu_\ell$.
The structure of the $(tbW)$ vertex has been probed in this way in a number of
experimental investigations~\cite{Aaboud:2017aqp,Khachatryan:2014vma,%
Aad:2015yem,Khachatryan:2016sib}. It is clear that, in a perturbative
next-to-leading order (NLO) calculation, one has to complement the (Born
$\otimes$ one-loop) contributions to the spin density matrix by the integrated
(tree $\otimes$ tree) contributions. In the second approach one considers the
quasi-three-body decay $t(\uparrow) \to X_b + \ell^+ + \nu_\ell$ which is
analyzed entirely in the top quark rest frame. 

The general matrix element for the decay $t \to b+W^+$ including the
leading-order (LO) standard model (SM) contribution is written
as~\cite{Bernreuther:1992be,Bernreuther:2008us,AguilarSaavedra:2006fy,%
GonzalezSprinberg:2011kx} 
\be\label{genme}
M_\mu (tbW^+)=-\frac{g_W}{\sqrt{2}} \bar u_b
\Big[\gamma_\mu ((V_{tb}^\ast +f_L) P_L+f_R P_R) 
  +\frac{i\sigma_{\mu\nu}\,q^\nu}{m_W}\,\Big(g_L P_L+g_R P_R\Big)\Big]u_t
\en
where $P_{L,R}=(1\mp\gamma_5)/2$. The LO SM structure of the $(tbW)$ vertex is
obtained by dropping all terms except for the contribution proportional to
$V^\ast_{tb}\sim 1$. The form factors are in general complex-valued functions
where SM imaginary parts can be generated by $CP$-conserving final state
interactions or can be introduced by hand as non-SM $CP$-violating imaginary
contributions. 

The set of observables in polarized top quark decays divide into two classes
-- the $T$-even and $T$-odd observables. The $T$-even observables, including
their NLO QCD corrections have been discussed before in
Refs.~\cite{Fischer:1998gsa,Fischer:2001gp} (sequential decays) and in
Refs.~\cite{Czarnecki:1990pe,Czarnecki:1993gt,Czarnecki:1994pu,Groote:2006kq}
(quasi-three-body decays). This paper is devoted to a detailed analysis of the
$T$-odd observables contributing to polarized top quark decays. These are
either fed by $CP$-conserving SM final state interactions or by $CP$-violating
non-SM interactions.   

The matrix element~(\ref{genme}) is folded with the Born term contribution to
obtain the relevant $T$-odd contributions. In the case $m_b=0$ (which we use
throughout the paper) it turns out that only the coefficient $\imag g_R$
generates $T$-odd correlations. $T$-odd correlations can be studied in both
the sequential decay $t(\uparrow) \to X_b +W^+ (\to \ell^+ +\nu_\ell)$ and the
quasi-three-body decay $t(\uparrow) \to X_b+ \ell^+ + \nu_\ell$. In either
case we count the number of $T$-odd observables, determine the angular factors
that multiply them in the relevant angular decay distributions and quantify
them in terms of the contribution of the imaginary part of the right-chiral
tensor coupling $g_R$.

We discuss the two approaches in turn in Secs.~\ref{sec2} and~\ref{sec3}
where we concentrate on the $T$-odd contributions to these decays. We comment
on the relations between the two approaches at the end of Sec.~\ref{sec3}. In
Sec.~\ref{sec4} we discuss positivity constraints on the various coupling
factors in Eq.~(\ref{genme}) resulting from the requirement that the
differential angular decay rate has to remain positive definite over the full
angular phase space.

In Sec.~\ref{sec5} we discuss the electroweak contributions to $\imag g_R$.
Contrary to the case of QCD the NLO electroweak corrections admit absorptive
one-loop vertex contributions, or, put in a different language, of final state
interactions/rescattering corrections. The absorptive parts of the NLO
electroweak one-loop vertex contributions treated in this paper in the case
$m_b=0$ provide imaginary contributions to the coupling terms $f_L$ and $g_R$
where $\imag f_L$ does not contribute to the $T$-odd correlations. The reason
is that $f_L$ multiplies the same coupling structure as the Born term. The
results on $\imag g_R$ are presented in analytical form. The absorptive
contributions to $g_R$ have been calculated before analytically (for photon
exchange) and numerically (for $Z$ exchange) in
Refs.~\cite{GonzalezSprinberg:2011kx} and~\cite{Arhrib:2016vts}. We agree with
the results of Ref.~\cite{Arhrib:2016vts} up to small numerical differences
but disagree with the result of Ref.~\cite{GonzalezSprinberg:2011kx} for the
$Z$ exchange contribution. Finally, in Sec.~\ref{sec6} we provide a summary
of our results and present our conclusions.

\section{\label{sec2}Quasi-two-body decay $t(\uparrow) \to X_b+W^+$ followed\\
  by the decay $W^+ \to \ell^+ + \nu_\ell$ (sequential decay)}
Let us begin by counting the number of independent structure functions that
appear in the description of the sequential decay
$t(\uparrow) \to X_b+W^+(\to \ell^+ + \nu_\ell)$. This is best done by
considering the independent spin density matrix elements
$H_{\lambda_W\, \lambda'_W}^{\,\lambda^{\phantom x}_t\,\lambda'_t}$ of the
$W^+$ (also called helicity structure functions or, for short, structure
functions) which form a hermitian $(3\times3)$ matrix 
\be
\bigg(H_{\lambda_W\, \lambda'_W}^{\,\lambda^{\phantom x}_t\,\lambda'_t}
  \bigg)^\dagger=
\bigg(H_{\lambda'_W\, \lambda_W}^{\,\lambda'_t\,\lambda_t}\bigg)
\en
Since the spin of the $X_b$ state remains unobserved, one has the angular
momentum constraint $\lambda_t+\lambda_W=\lambda'_t+\lambda'_W$ implying
$|\lambda_W-\lambda'_W| \le 1$.
With the above constraints one counts ten independent double spin density
matrix elements:
\be
H_{++}^{++},\,H_{++}^{--},\,H_{--}^{++},\,H_{--}^{--},\,H_{00}^{++},\,%
H_{00}^{--},\,\real H_{+0}^{+-},\,\imag H_{+0}^{+-},\,\real H_{-0}^{-+},
\,\imag H_{-0}^{-+}.
\en
The two structure functions $\imag H_{+0}^{+-}$ and $\imag H_{-0}^{+-}$ are
so-called $T$-odd structure functions, the terminology of which will be
explained later on.

\begin{figure}\begin{center}
\epsfig{figure=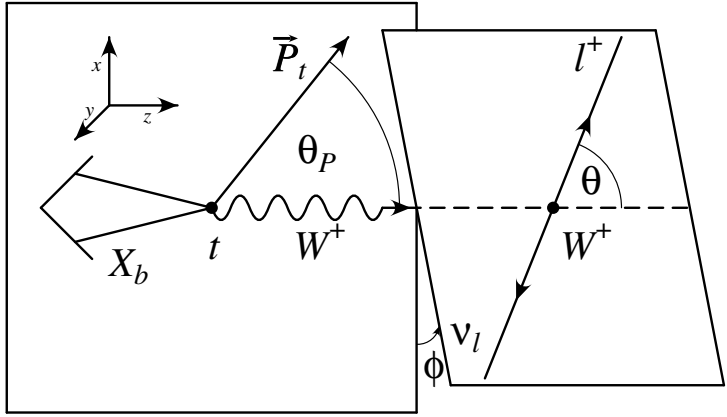, scale=0.55} 
\end{center}
\caption{\label{angdef} Definition of the polar angles $\theta$ and $\theta_P$
and the azimuthal angle $\phi$ in the sequential decay
$t(\uparrow) \to X_b +W^+ (\to \ell^+ +\nu_\ell)$}
\end{figure}

In the narrow width approximation the decay
$t(\uparrow) \to X_b+  \ell^+ +\nu_\ell$ can be described by a sequential
two-step decay process given by the  decays 
$t(\uparrow) \to X_b +W^+$ and $W^+ \to \ell^+ +\nu_\ell$. Accordingly one
defines two coordinate systems -- the top quark rest frame and the
$W^+$ rest frame -- where the repective angles in the two systems are
defined in Fig.~\ref{angdef}.

The $W^+$ produced in the decay $t(\uparrow) \to X_b +W^+$ is highly polarized.
The polarization of the $W^+$ can be analyzed in the angular decay distribution
of the decay $W^+ \to \ell^+ +\nu_\ell$. The full three-fold angular decay
distribution is obtained from the trace of the product of the spin-1 density
matrix of the $W^+$ in the production process $t \to b +W^+$ and the transpose
of the spin-1 density matrix describing the decay process
$W^+ \to \ell^+ + \nu_\ell$.

The production spin density matrix ${\cal H}_{\lambda_W\lambda'_W}(\theta_P)$
reads
\be\label{had}
{\cal H}_{\lambda_W\lambda'_W}(\theta_P) = 
\left( \begin{array}{ccc}
 H_{++}+ H_{++}^P P\cos\theta_P & H_{+0}^P\, P\sin\theta_P & 0\\
   H_{0+}^P P \sin\theta_P &  H_{00}+ H_{00}^P P\cos\theta_P 
& H_{0-}^P\, P \sin\theta_P \\ 
0 &  H_{-0}^P\, P \sin\theta_P  &  H_{--}+ H_{--}^P P\cos\theta_P  
\end{array}\right)
\en
In practice one works with a normalized spin density matrix
$\widehat {\cal H}_{\lambda_W\lambda'_W}={\cal H}_{\lambda_W\lambda'_W}/{\cal H}_{\rm tot}$
where ${\cal H}_{\rm tot}=H_{++}+H_{00}+H_{--}$. In addition, it is also useful
to extract the unit matrix $\oone$ from the normalized spin density
matrix (see e.g. Ref.~\cite{Aguilar-Saavedra:2015yza}).

The polarization of the top quark in the top quark rest frame is given by (see
Fig.~\ref{angdef})
\be\label{polvec}
\vec P_t= P\,(\sin\theta_P,0,\cos\theta_P)
\en
where $P$ is the magnitude of the polarization of the top quark. The relevant
helicity structure functions can be projected with the help of the spin-1
polarization four-vectors of the $W^+$ which, in the top quark rest frame,
is given by
\be
\varepsilon(0)^\mu =\frac{1}{\sqrt{q^2}}(|\vec q|;\,0,\,0,\,q_0) \qquad
\varepsilon(\pm)^\mu= \frac{1}{\sqrt{2}}(0;\,\mp 1,\,-i,\,0)
\en
The longitudinal and transverse polarization components of the top quark are
given by $s_t^{\ell\,\mu}=(0;0,0,1)$ and $s_t^{tr\,\mu}=(0;1,0,0)$, again in
the top quark rest frame. The diagonal elements ($\lambda_W = \lambda'_W$) of
${\cal H}_{\lambda_W\lambda'_W}$ are defined  by
\bea {\rm diagonal\,\,unpolarized}\qquad
{\cal H}_{\lambda_W\lambda_W}&=&{\cal H}_{\mu\,\nu}
 \varepsilon^{\ast\,\mu}(\lambda_W)  \varepsilon^{\nu}(\lambda_W) \nn
{\rm diagonal\,\, polarized}\qquad {\cal H}^P_{\lambda_W\lambda_W}
&=&{\cal H}_{\mu\,\nu}(s_t^\ell)
 \varepsilon^{\ast\,\mu}(\lambda_W)  \varepsilon^{\nu}(\lambda_W)
\ena
while the off-diagonal polarized elements ($\lambda_W \neq \lambda'_W$) are
determined by
\be
{\cal H}^P_{\lambda_W\lambda'_W}
={\cal H}_{\mu\,\nu}(s_t^{tr})
 \varepsilon^{\ast\,\mu}(\lambda_W)  \varepsilon^{\nu}(\lambda'_W) 
\en
Again, from angular momentum conservation one has
$\lambda_W - \lambda'_W = \pm 1 = \lambda'_t- \lambda_t$. The two
configurations $(\lambda_t\,,\lambda'_t) = ( 1/2,-1/2),\, (-1/2, 1/2)$ are
associated with the transverse polarization of the top quark (for details
see Ref.~\cite{Fischer:2001gp}).

The leptonic spin density matrix ${\cal L}_{\lambda_W\lambda'_W}$ can be
projected in a similar way. One obtains (see e.g.\ Ref.~\cite{Gutsche:2015mxa})
\bea\label{lep}
 &&\!\!\!{\cal L}_{\lambda_W\lambda'_W}(\theta,\phi) = \frac{q^2}{2}\times \nn
&&\left( \begin{array}{ccc}
  (1+\cos\theta)^2 & \frac{2}{\sqrt{2}} (1+\cos\theta)\sin\theta\, e^{i\phi}
  & \sin^2\theta\, e^{2i\phi} \\
  \frac{2}{\sqrt{2}} (1+\cos\theta)\sin\theta\, e^{-i\phi}
  & 2\sin^2\theta 
&  \frac{2}{\sqrt{2}} (1-\cos\theta)\sin\theta\, e^{i\phi} \\
 \sin^2\theta e^{-2i\phi} 
&  \frac{2}{\sqrt{2}} (1-\cos\theta)\sin\theta\, e^{-i\phi} & (1-\cos\theta)^2 
\end{array} \right)
\ena
where the angles $\theta$ and $\phi$ are defined in Fig.~\ref{angdef}. We have
set $m_\ell=0$ in Eq.~(\ref{lep}). The angular decay distribution is then
obtained from~\cite{Fischer:2001gp,Aguilar-Saavedra:2015yza,Bialas:1992ny,%
Aguilar-Saavedra:2017wpl,Boudreau:2013yna,Mueller:2016}
\be\label{angdist}
W(\theta,\theta_P,\phi)=
  \sum_{\lambda_W \lambda'_W} {\cal H}_{\lambda_W\lambda'_W}(\theta_P)
  {\cal L}_{\lambda_W\lambda'_W}(\theta,\phi)
  =\Tr \left ({\cal H}(\theta_P) \cdot
{\cal L}^T(\theta,\phi) \right)
\en
Here we concentrate on the $T$-odd correlations in the angular decay
distribution~(\ref{angdist}). The $T$-odd pieces are given
by the terms in Eq.~(\ref{angdist}) proportional to $\sin\phi$. One has 
\bea\label{t-odd}
W^{T{\rm-odd}}(\theta,\theta_P,\phi)&=&q^2\Big(-\sqrt{2} H_{{\cal I}I}^P
\sin\theta_P\,\sin2\theta\,\sin\phi \nn
&&+2 \sqrt{2} H_{{\cal I}A}^P\sin\theta_P\,\sin\theta\,\sin\phi \Big)
\ena
where we define two $T$-odd helicity structure functions
by~\cite{Groote:1995ky,Groote:1996nc}
\bea
H_{{\cal I}I}^P&=&\frac{-i}{4}\Big(H^P_{+0}-H^P_{0+}+H^P_{-0}-H^P_{0-}\Big)\nn
H_{{\cal I}A}^P&=&\frac{-i}{4}\Big(H^P_{+0}-H^P_{0+}-H^P_{-0}+H^P_{0-}\Big)
\ena
Compared to Refs.~\cite{Groote:1995ky,Groote:1996nc} we have changed the
notation for the $T$-odd structure functions such that
$(H_5,\,H_9)\, \to\, (H_{{\cal I}I},\,H_{{\cal I}A}$).

That the two angular factors in Eq.~(\ref{t-odd}) correspond to $T$-odd
correlations can be seen by representing the angular factors in
Eq.~(\ref{t-odd}) in terms of triple-products. To demonstrate this we collect
the relevant normalized three-vectors as defined in Fig.~\ref{angdef}. They
read
\be
\hat p_\ell=(\sin\theta\,\cos\phi,\sin\theta\,\sin\phi,\cos\theta)
\qquad \hat q=(0,0,1) \qquad \hat P_t=(\sin\theta_P,0,\cos\theta_P)
\en
One then finds
\bea
\sin\theta_P\,\sin\theta\,\sin\phi&=&
\hat q\cdot(\hat P_t \times \hat p_{\ell})
\nn
\sin\theta_P\,\sin2\theta\,\sin\phi&=&2\,(\hat p_{\ell}\cdot \,\hat q)\,\,
\hat q\cdot(\hat P_t \times \hat p_{\ell})
\label{t-odd2}
\ena

The nomenclature $T$-odd interaction derives from the fact that a product
consisting of an odd number of momenta or polarization vectors as
in Eq.~(\ref{t-odd2}) changes sign under the time-reversal operation $t\to -t$
since the three-momentum $\vec p$ and the  polarization vector $\vec P_t$
transform as $(\vec p,\vec P_t) \to (-\vec p,-\vec P_t)$ under $t \to -t$.

After having set up the general formalism, we are now ready to discuss the
contribution of $\imag g_R$ to the $T$-odd structure functions. We shall work
at the leading level, i.e.\ we take the final state to be made up of a single
bottom quark and a $W^+$. That is, we now deal with $t \to b+ W^+$ instead of
$t \to X_b+W^+$. We also treat the contributions of $f_L$, $f_R$, $g_L$ and
$g_R$ as small perturbations. We thus keep only terms linear in $f_L$, $f_R$,
$g_L$ and $g_R$ when we fold these with the SM Born term.

We further assume $m_b=0$. In the case $m_b = 0$, there are a number of
simplifications. For once, in the linear approximation there are no
interference terms between the left-chiral Born term and the right-chiral
coupling terms $f_R$ and $g_L$. This implies that the massless bottom quark
contributes effectively only with its negative helicity state, i.e.\
$\lambda_b=-1/2$. This implies that $\lambda_W \neq 1$ due to the angular
momentum constraint $\lambda_t=\lambda_W -\lambda_b$. It follows that the
four density matrix elements $H^P_{+0}$, $H^P_{0+}$, $H_{++}$ and $H^P_{++}$
vanish, i.e.\ the hadronic double spin density matrix
${\cal H}_{\lambda_W\lambda'_W}(\theta_P)$ reduces to a $2\times2$ matrix. In
particular, this means that the two independent $T$-odd observables in the
sequential decay $t(\uparrow) \to X_b+W^+(\to \ell^+ + \nu_\ell)$ coalesce to
a single observable.

For $m_b=0$ one is effectively dealing only with two complex-valued invariant
form factors in the decomposition of Eq.~(\ref{genme}). These are the form
factors $f_L$ and $g_R$. The number of independent invariant amplitudes agrees
with the number of independent helicity amplitudes to which they are linearly
related. The two independent helicity amplitudes are 
$H_{\lambda_b\,\lambda_W}=H_{-1/2\,0}\,,\,H_{-1/2\,-1}$.

Next we calculate the contribution of $\imag g_R$ to the structure functions
$H^P_{{\cal I}I}=-H^P_{{\cal I}A}$. The calculation can be streamlined by
making use of an interesting insight provided some time ago by
Kuruma~\cite{Kuruma:1992if}. For $m_b=0$ the longitudinal and transverse
projections of the matrix element~(\ref{genme}) are proportional to the
corresponding projections of the Born term matrix
elements~\cite{Kuruma:1992if}. In fact, using the covariant representation
of the longitudinal polarization four-vector
\be
\varepsilon^\mu(0)= -\,\frac{q^2 p_t^\mu - p_tq \,q^\mu}{\sqrt{q^2}
  \sqrt{(p_t q)^2-q^2 m_t^2}}
\en
it is not difficult to see that ($x=m_W/m_t$)
\be\label{lproject}
\varepsilon^{\ast\,\mu}(0) M_\mu=\Big(1+f_L -x\, g_R\Big)\,\,
  \varepsilon^{\ast\,\mu} (0) M_\mu({\it Born})
\en
For the transverse projection one similarly finds  
\be\label{tproject}
\varepsilon^{\ast\,\mu}(-) M_\mu=\Big(1+f_L-\frac{1}{x}\,\,g_R\Big)\,\,
  \varepsilon^{\ast\,\mu} (-) M_\mu({\it Born})
\en
where the derivation of the factorization property is facilitated by making use
of the Gordon-type identity
\be
\bar u_b \frac{i\sigma_{\mu\nu}q^\nu}{m_W}P_R u_t =\bar u_b \left(-\frac{1}{x}
  \gamma_\mu P_L + \frac{1}{x}\,\frac{(2p_{t\,\mu}-q_\mu)}{m_t}P_R \right)u_t
\en

To proceed we calculate the Born term spin density matrix elements of the
$W^+$ needed when using Eqs.~(\ref{lproject}) and~(\ref{tproject}). The
corresponding Born term decay tensor $B^{\mu\,\nu}$ reads
\bea
B^{\mu\,\nu}=\sum_{\rm spins}M^\mu({\it Born})M^{\dagger\,\nu}({\it Born})
  &=& \tr \Big\{ \slp_b \gamma^\mu \tfrac 12(1-\gamma_5)
    (\slp_t+m_t)(1+\gamma_5 \sls_t)\gamma^\nu \tfrac 12(1-\gamma_5)\Big\} \nn
  &=& 2\left(\bar p_t^\mu p_{b}^\nu + \bar p_t^\nu p_{b}^\mu
     -\bar p_t \cdot p_{b} g^{\mu\nu} - i\epsilon^{\mu\nu\alpha\beta}
     \bar p_{t\,\alpha}
     p_{b\,\beta}\right)
\ena
where $\bar p_t=p_t - m_t s_t$. The Born term spin density elements $B_i$ and
$B_i^P$ have been listed in Ref.~\cite{Fischer:2001gp}. For the nondiagonal
structure functions discussed here one has to specify
$s_t^\mu=s_t^{tr \mu} = (0;1,0,0)$ (see Fig.~\ref{angdef}). One has 
\be
B^P_I=-B^P_A= -\frac 14(B_{-0}+B_{0-})
  = -\frac 12 \sqrt{2} m_t^2 \frac{1-x^2}{x}
\en

As discussed before we keep only terms linear in $f_L$ and $g_R$ when
calculating the structure functions $H^P_{-0}$ and $H^P_{0-}$ assuming that
the form factors are small. One has
\bea
H^P_{-0}&=&\Big(1+ 2\real f_L - \frac{1+x^2}{x}\real g_R
  - i \frac{1-x^2}{x}\imag g_R \Big) B^P_{-0}\nn
H^P_{0-}&=&\Big(1+ 2\real f_L - \frac{1+x^2}{x}\real g_R
  + i \frac{1-x^2}{x}\imag g_R \Big) B^P_{0-}
\ena
The NLO imaginary contribution $\imag f_L$ does not contribute to the
nondiagonal matrix elements $H^P_{-0}=H^{P\ast}_{0-}$ because the matrix
element $f_L$ multiplies the same covariant $\gamma_\mu P_L$ as the Born term;
i.e., it is self-interfering.

For the $T$-odd structure functions one finally obtains
\bea
{\cal H}^P_{{\cal I}I} =-{\cal H}^P_{{\cal I}A}
  &=& - \frac{i}{4}\left(H^P_{-0} - H^P_{0-}\right)
  \ =\ \frac{1-x^2}{x}\,\imag g_R\, B^P_{I} \nn
  &=& - \frac{m_t^2}{\sqrt{2}} \left(\frac{1-x^2}{x}\right)^2 \imag g_R
\ena
The $m_b=0$ $T$-odd angular decay distribution reads
\bea\label{t-odd3}
W^{T{\rm-odd}}(\theta,\theta_P,\phi)&=&2 \sqrt{2}\,{\cal  H}^P_{{\cal I}I}\,
  P_t \sin\theta_P\,\sin\theta\,(1-\cos\theta)\,\sin\phi \nn
  &=&-2 m_t^4 (1-x^2)^2 \imag g_R
  P_t \sin\theta_P\,\sin\theta\,(1-\cos\theta)\,\sin\phi
\ena
with an overall factor $(1-\cos\theta)$ as expected from angular momentum
conservation.

In order to get a feeling for the size of the $T$-odd contribution relative to
the unpolarized rate we integrate the full angular decay distribution over
$\cos\theta$ where we keep only the Born term contributions in the $T$-even
terms. One has
\bea\label{angdist2}
W(\theta_P,\phi)&=&\int_{-1}^{1} d\cos\theta \,\,W(\theta,\theta_P,\phi) \nn
  &=&\frac 43 m_t^4(1-x^2)(1+2x^2)\Bigg(1+\frac{(1-2x^2)}{(1+2x^2)}\cos\theta_P
  \nn
&&+\frac 34 \pi \frac{x}{(1+2x^2)} \sin\theta_P \cos\phi
  -\frac 34 \pi \frac{(1-x^2)}{(1+2x^2)}\imag g_R
  P_t \sin\theta_P\,\sin\phi\Bigg)
\ena
The factor $3\pi(1-x^2)/(4(1+2x^2))=1.29$ multiplying $\imag g_R$ is
sufficiently large to make an angular analysis such as Eq.~(\ref{angdist2})
promising.

\section{\label{sec3} Quasi-three-body decays
  $t(\uparrow) \to X_b + \ell^+ + \nu_\ell$}
In this variant of possible angular decay distributions the decay
$t(\uparrow) \to X_b + \ell^+ + \nu_\ell$ is analyzed entirely in the top quark
rest frame. Let us begin again by enumerating the number of structure functions
that appear in the quasi-three-body decay
$t(\uparrow) \to X_b + \ell^+ + \nu_\ell$. These are the two complex matrix
elements $M_{\lambda_t=1/2}$ and $M_{\lambda_t=-1/2}$ that describe the
transition $t(\uparrow) \to X_b + \ell^+ + \nu_\ell$. One thus has altogether
the four structure functions $|M_{1/2}|^2$, $|M_{-1/2}|^2$,
$\real M_{1/2}M^*_{-1/2}$ and $\imag M_{1/2}M^*_{-1/2}$ needed to represent
the decay process.

\begin{figure}
\begin{center}
\epsfig{figure=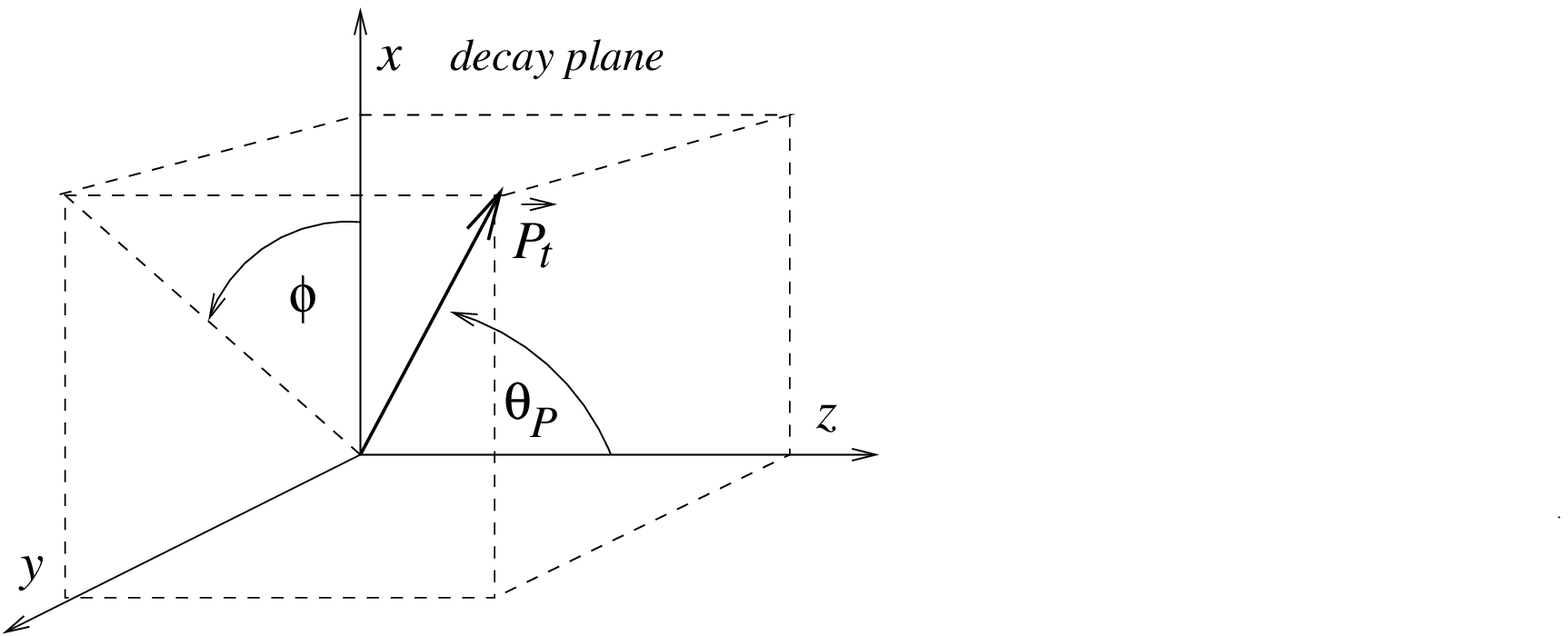, scale=0.8}\kern-164pt
\epsfig{figure=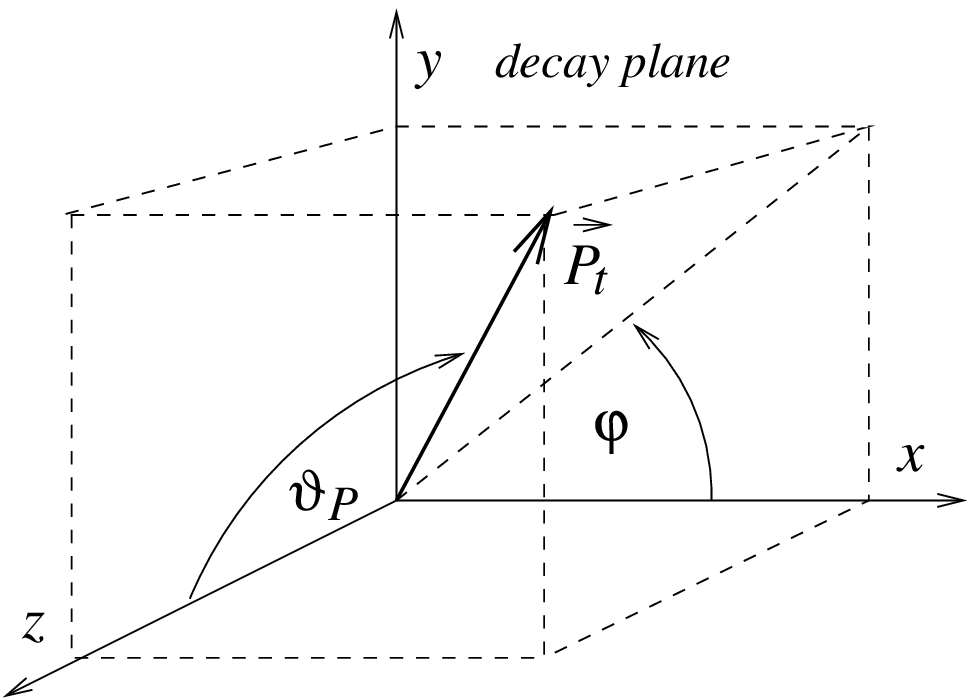, scale=0.8}\\
(a)\kern220pt(b) 
\end{center}
\caption{\label{systemdef} Definition of the helicity system (left panel (a))
  and the transversity system (right panel (b)) in the quasi-three-body
  decay $t(\uparrow) \to X_b + \ell^+ + \nu_\ell$. The polar angles $\theta_P$
  and $\vartheta_P$ and the azimuthal angles $\phi$ and $\varphi$ describe
  the orientation of the polarization vector $\vec P$ of the top quark in the
  two systems.}
\end{figure}

The angular decay distribution of the decay is obtained by folding the decay
matrix $M_{\lambda_t}M^*_{\lambda'_t}$ with the spin density matrix of the top
quark, i.e.\ by calculating the trace
$\Tr(\rho_{\lambda_t\,\lambda'_t}\,M_{\lambda_t}M^*_{\lambda'_t})$ where the
spin density matrix of the top quark is given by
\be
\rho_{\lambda_t\,\lambda'_t} = \oone + P_{t\,z}\,\,\ssigma_z
+P_{t\,x}\,\,\ssigma_x +P_{t\,y}\,\,\ssigma_y
\en
($\ssigma_i$, $i=x,y,z$ are the Pauli matrices). The components of the
polarization vector $\vec P_t=(P_{t\,i})$ depend on the coordinate system in
which the decay is analyzed. There is a multitude of possible choices for the
decay coordinate system. Two different classes of coordinate systems have been
in use in the literature -- the helicity system and the transversity system.
In the helicity system the three final state momenta in the top quark rest
frame span the $(x,z)$ plane while, in the transversity system, they span the
$(x,y)$ plane. The two classes of systems are displayed in Fig.~\ref{systemdef}
together with the definition of the respective polar and azimuthal angles
describing the orientation of the polarization vector of the top quark.
Depending on the choice of coordinate system the polarized structure functions
get toggled around among the various angular factors that multiply them. We
shall discuss these two possible choices in turn. Which of the systems are
being used in the experimental analysis has to be decided on the experimental
expediency.

\subsection{The helicity system}
\begin{figure}\begin{center}
\epsfig{figure=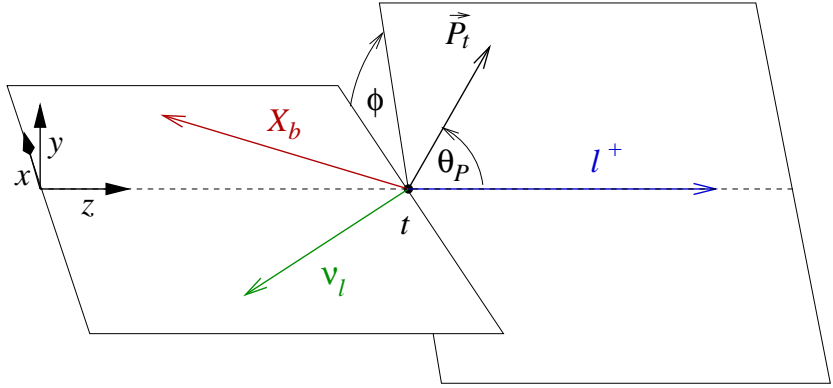, scale=0.55} 
\end{center}
\caption{\label{Ib} Definition of the polar angles $\theta_P$ and the azimuthal
angle $\phi$ in the helicity system Ib for the quasi-three-body decay
$t(\uparrow) \to X_b + \ell^+ +\nu_\ell$}
\end{figure}

In the following, we limit our attention to three helicity systems where the
decay plane is in the $(x,z)$ plane and the $z$-axis points into the $\ell^+$
direction, the $X_b$ direction or the $\nu_\ell$ direction. One further has to
specify the orientation of the $x$ axis relative to the event. We thus define
six coordinate systems according to
\bea
\label{sysclass}
    {\rm system\,I}\,&:& \vec p_\ell \parallel z\,;\qquad a\,:
    \,p_{\nu\,x}\ge0\qquad b\,:\,p_{X_b\,x}\ge 0 \nn
    {\rm system\,II}\,&:& \vec p_{X_b} \parallel z\,;\qquad a\,:
    \,p_{\ell\,x}\ge 0\qquad b\,:\,p_{\nu\,x}\ge 0 \nn
    {\rm system\,III}\,&:& \vec p_\nu \parallel z\,;\qquad a\,:
    \,p_{X_b\,x}\ge0\qquad b\,:\,p_{\ell\,x}\ge 0 
\ena
When labelling the three systems we follow the conventions of
Ref.~\cite{Korner:1998nc}. For instance, in system Ib the momenta and
polarization vector read~\cite{Groote:2006kq} (see Fig.~\ref{Ib})
\bea\label{momhel}
p_t &=& m_t(1;0,0,0) \nn
p_\ell &=& \frac{m_t}{2}x_\ell\,(1;0,0,1) \nn
p_\nu &=& \frac{m_t}{2}(1-x_\ell +x^2)(1;-\sin\theta_\nu,0,\cos\theta_\nu) \nn
p_b &=& \frac{m_t}{2}(1-x^2)(1;\sin\theta_b,0,\cos\theta_b) \nn
s_t &=& (0,\sin\theta_P \cos\phi,\sin\theta_P \sin\phi,\cos\theta_P)
\ena
where $x_\ell=2E_\ell/m_t$ is the scaled lepton energy and
\bea
\cos\theta_\nu&=&\frac{x_\ell(1-x_\ell+x^2)-2x^2}{x_\ell(1-x_\ell+x^2)}\qquad
\sin\theta_\nu\ =\ \frac{2x\sqrt{(1-x_\ell)(x_\ell-x^2)}}{x_\ell(1-x_\ell+x^2)}
  \nn
\cos\theta_b&=&\frac{2x^2-x_\ell(1+x^2)}{x_\ell(1-x^2)}\qquad
\sin\theta_b=\frac{2x}{x_\ell(1-x^2)}\sqrt{(1-x_\ell)(x_\ell-x^2)}
\ena

For the spin density matrix of the top quark one has
\be
\rho_{\lambda_t\,\lambda'_t} = \oone + P_t \cos\theta_P\,\ssigma_z
  +P_t\sin\theta_P\cos\phi\,\ssigma_x+P_t\sin\theta_P\sin \phi\,\ssigma_y
\en
where $\theta_P$ and $\phi$ describe the orientation of the polarization vector
 of the top quark as can be read off from Fig.~\ref{systemdef}(a). We expand
the $(2\times2)$ decay matrix $M_{\lambda_t}M^*_{\lambda'_t}$ along the unit
matrix $\oone$ and the three $\ssigma_i$ matrices. One has
\be
M_{\lambda_t}M^*_{\lambda'_t}= \tfrac 12 \left(A\,\oone+ B\,\ssigma_z
  +C\,\ssigma_x +D\,\ssigma_y\right)
\en
The angular decay distribution of the decay is obtained by folding the decay
matrix $M_{\lambda_t}M^*_{\lambda'_t}$ with the spin density matrix of
the top quark, i.e.\ by calculating the trace
$\Tr(\rho_{\lambda_t\,\lambda'_t}\,M_{\lambda_t}M^*_{\lambda'_t})$. One obtains
\bea\label{angdisthel}
W(\theta_P,\phi) &=&\Tr\Big\{\rho_{\lambda_t\,\lambda'_t}
  M_{\lambda_t}M^*_{\lambda_t'}\Big\} \nn
  &=& A + B\ P_t\cos\theta_P + C\ P_t \sin\theta_P\cos\phi
  + D\ P_t\sin\theta_P\sin\phi 
\ena
The term proportional to the structure function $D$ represents the $T$-odd
contribution as can be seen by the representation
\be
\sin\theta_P \sin\phi =\frac{1}{\sin\theta_\nu}\,\hat p_\nu\cdot
  (\hat p_\ell \times \hat s_t)
\en
        
The structure functions $A(x_\ell)$, $B(x_\ell)$, $C(x_\ell)$ and $D(x_\ell)$
can be calculated from the contraction of the hadron and lepton tensors given
by ${\cal H}^{\mu\nu}{\cal L}_{\mu\nu}$. Including the LO contribution
proportional to $V_{tb}^\ast\sim 1$ one obtains for the six different systems
\bea
\label{helsysunin}
\lefteqn{({\cal H}^{\mu\nu}{\cal L}_{\mu\nu})_{\rm I\,a/b}
  \ =\ 4m_t^4(1-x_\ell)\Big[\left(x_\ell(1+2\real f_L)-2x\real g_R\right)
  (1+P_t\cos\theta_P)\strut}\nonumber\\&&\strut\qquad
  \pm2\sqrt{(1-x_\ell)(x_\ell-x^2)}\real g_RP_t\sin\theta_P\cos\phi
  \strut\nonumber\\&&\strut\qquad
  \mp2\sqrt{(1-x_\ell)(x_\ell-x^2)}\imag g_RP_t\sin\theta_P\sin\phi\Big]
  \nonumber\\[12pt]
   \lefteqn{({\cal H}^{\mu\nu}{\cal L}_{\mu\nu})_{\rm II\,a/b}
  \ =\ \frac{4m_t^4(1-x_\ell)}{1-x^2}\Big[(1-x^2)(x_\ell(1+2\real f_L)
  -2x\real g_R)\strut}\nonumber\\&&\strut\qquad
  -\left(((1+x^2)x_\ell-2x^2)(1+2\real f_L)+2x(1+x^2-2x_\ell)\real g_R\right)
  P_t\cos\theta_P\strut\nonumber\\&&\strut\qquad
  \pm\sqrt{(1-x_\ell)(x_\ell-x^2)}\left(2x(1+2\real f_L)
  -2(1+x^2)\real g_R\right)P_t\sin\theta_P\cos\phi\strut\nonumber\\
  &&\strut\qquad
  \mp2(1-x^2)\sqrt{(1-x_\ell)(x_\ell-x^2)}\imag g_R
  P_t\sin\theta_P\sin\phi\Big]\nonumber\\[12pt]
\lefteqn{({\cal H}^{\mu\nu}{\cal L}_{\mu\nu})_{\rm III\,a/b}
  \ =\ \frac{4m_t^4(1-x_\ell)}{1-x_\ell+x^2}\Big[(1-x_\ell+x^2)
  \left(x_\ell(1+2\real f_L)-2x\real g_R\right)\strut}\nonumber\\&&\strut\qquad
  +\left(\left((1-x_\ell+x^2)x_\ell-2x^2\right)(1+2\real f_L)
  +2x(1-x_\ell+x^2)\real g_R\right)P_t\cos\theta_P
  \strut\nonumber\\&&\strut\qquad
  \mp\sqrt{(1-x_\ell)(x_\ell-x^2)}\left(2x(1+2\real f_L)
  -2(1-x_\ell+x^2)\real g_R\right)P_t\sin\theta_P\cos\phi
  \strut\nonumber\\&&\strut\qquad
  \mp2(1-x_\ell+x^2)\sqrt{(1-x_\ell)(x_\ell-x^2)}\imag g_R
  P_t\sin\theta_P\sin\phi\Big].
\ena
After integration over $x_\ell$ in the limits $x^2\le x_\ell \le 1$ one obtains
\begin{eqnarray}\label{helsysin}
\int dx_\ell\,\,\lefteqn{({\cal H}^{\mu\nu}{\cal L}_{\mu\nu})_{\rm I\,a/b}
  \ =\ \frac{(1-x^2)^2 m_t^4}6\Big[4\left((1+2x^2)(1+2\real f_L)
  -6x\real g_R\right)(1+P_t\cos\theta_P)\strut}\nonumber\\&&\strut
  \pm 3\pi(1-x^2)\real g_RP_t\sin\theta_P\cos\phi
  \mp 3\pi(1-x^2)\imag g_RP_t\sin\theta_P\sin\phi\Big]\nonumber\\[12pt]
\int dx_\ell\,\,\lefteqn{({\cal H}^{\mu\nu}{\cal L}_{\mu\nu})_{\rm II\,a/b}
  \ =\ \frac{(1-x^2)^2 m_t^4}6\Big[4\left((1+2x^2)(1+2\real f_L)
  -6x\real g_R\right)\strut}\nonumber\\&&\strut\qquad
  -4\left((1-2x^2)(1+2\real f_L)+2x\real g_R\right)P_t\cos\theta_P
  \strut\nonumber\\&&\strut
  \pm3\pi\left(x(1+2\real f_L)-(1+x^2)\real g_R\right)
  P_t\sin\theta_P\cos\phi\strut\nonumber\\&&\strut
  \mp3\pi(1-x^2)\imag g_RP_t\sin\theta_P\sin\phi\Big]\nonumber\\[12pt]
\int dx_\ell\,\,\lefteqn{({\cal H}^{\mu\nu}{\cal L}_{\mu\nu})_{\rm III\,a/b}
  \ =\ \frac{m_t^4}6\Big[4(1-x^2)^2\left((1+2x^2)(1+2\real f_L)
  -6x\real g_R\right)\strut}\nonumber\\&&\strut
  +4\Big(\left((1-x^2)(1-11x^2-2x^4)-24x^4\ln x\right)(1+2\real f_L)
  \strut\nonumber\\&&\strut\qquad
  +6x(1-x^2)^2\real g_R\Big)P_t\cos\theta_P\strut\nonumber\\&&\strut
  \mp 3\pi(1-x)^3\left(2x(1+3x)(1+2\real f_L)-(1+x)^3\real g_R\right)
  P_t\sin\theta_P\cos\phi\strut\nonumber\\&&\strut
  \mp 3\pi(1-x^2)^3\imag g_RP_t\sin\theta_P\sin\phi\Big]
\end{eqnarray}
A few comments on the structure of the various contributions are in order.
\begin{itemize}
\item After azimuthal averaging and dropping the non-SM contributions
$\real f_L$ and $\real g_R$ one obtains from Eq.~(\ref{helsysin}) the
well-known polar distributions
\be
W(\theta)=1+ \kappa_i P_t\cos\theta \qquad {\rm with} \qquad
\left\{ \begin{array}{l}
  \kappa_{\rm I}=1 \\
  \kappa_{\rm II} =(1-2x^2)/(1+2x^2)=0.398\\
  \kappa_{\rm III}= f(x)=-0.261
  \end{array} \right\}
\en
where
\be
f(x)=\frac{(1-x^2)(1-11x^2-2x^4)-24x^4\ln x}{(1-x^2)^2(1+2x^2)}
\en
\item The results of systems II and III can be obtained from the results of
system I through a rotation around the $y$ axis. The relevant rotations read
\be
\left(\begin{array}{c}
    B_{\rm II} \\
    C_{\rm II}
    \end{array} \right)=
  \left( \begin{array}{cc}
    \cos\theta_b & -\sin\theta_b \\
    \sin\theta_b & \cos\theta_b
  \end{array} \right)
  \left( \begin{array}{c}
    B_{\rm I} \\
    C_{\rm I}
    \end{array} \right)
  \en
  \be
  \left( \begin{array}{c}
    B_{\rm III} \\
    C_{\rm III}
    \end{array} \right)=
  \left( \begin{array}{cc}
    \cos\theta_\nu & \sin\theta_\nu \\
    -\sin\theta_\nu & \cos\theta_\nu
  \end{array} \right)
  \left( \begin{array}{c}
    B_{\rm I} \\
    C_{\rm I}
    \end{array} \right)
  \en
Since the unpolarized rate function $A$ and the $T$-odd $y$ component $D$ are
not affected by this rotation the structure functions $A$ and $D$ are the same
in all three systems such that e.g.\ $D_{\rm I}=D_{\rm II}=D_{\rm III}$.
\item The decay distribution in system IIa is closely related to the decay
distribution of sequential top quark decay discussed in Sec.~\ref{sec2}. In
fact, take Eq.~(\ref{angdist}) and substitute the relation between the cosine
of the angle $\theta$ and the scaled lepton energy $x_\ell$ for
$m_b=0$ (see e.g.\ Ref.~\cite{Kadeer:2005aq})
\be
\cos\theta=\frac{(x_\ell-x^2)-(1-x_\ell)}{(1-x^2)}\qquad
\sin\theta=\frac{2}{1-x^2}\sqrt{(x_\ell-x^2)(1-x_\ell)}
\en
into Eq.~(\ref{angdist}). One then recovers the unintegrated distribution
$({\cal H}^{\mu\nu}{\cal L}_{\mu\nu})_{\rm II\,a}$ after the replacement
$\theta_P\to \pi-\theta_P$. The structure functions describing the
quasi-three-body decays can be seen to be weighted sums of the unpolarized and
polarized helicity structure functions in the sequential decays with weight
functions $w(x_\ell)$ that are not simple. It is only the $T$-odd structure
functions that have a simple one-to-one relation. The relation between the
$T$-odd structure functions ${\cal H}_{{\cal I}A}^P$ and
$D_{\rm Ia}=D_{\rm IIa}=D_{\rm IIIa}=D_{\rm a}$ can be worked out to be
\be
D_{\rm a}=m_t^2 8 \sqrt{2}(1-x_\ell)\sqrt{(1-x_\ell)(x_\ell-x^2)}
\frac{x^2}{(1-x^2)}\, {\cal H}_{{\cal I}A}^P
\en
When comparing the corresponding expressions integrated over $\cos\theta$ and
$x_\ell$ one has to take into account the change in integration measure
$d\cos\theta/dx_\ell=2/(1-x^2)$.
\end{itemize}

\subsection{The transversity system}
The event plane is now in the $(x,y)$ plane and the $z$ axis is defined
by the normal to the event plane as shown in Fig.~\ref{systemdef}(b). The
angles in the helicity system and the transversity system are related by
\bea\label{anglerel}
\cos\vartheta_P&=&\sin\theta_P \sin\phi \nn
\sin\vartheta_P \sin\varphi&=& \sin\theta_P \cos\phi \nn
\sin\vartheta_P \cos\varphi&=& \cos\theta_P 
\ena
These relations can be obtained by geometric reasoning or, more directly, by
evaluating the scalar products $(p_\ell\cdot s_t)$, $(p_b\cdot s_t)$ and
$\varepsilon(p_t,p_\ell,p_b,s_t)$ in the two systems using the momentum
representation in helicity system Ib listed in Eq.~(\ref{momhel}) and the
corresponding representation in the transversity system
\bea
p_t &=& m_t(1;0,0,0) \nn
p_\ell &=& \frac{m_t}{2}x_\ell\,(1;1,0,0) \nn
p_\nu &=& \frac{m_t}{2}(1-x_\ell +x^2)(1;\cos\theta_\nu,-\sin\theta_\nu,0) \nn
p_b &=& \frac{m_t}{2}(1-x^2)(1;\cos\theta_b,\sin\theta_b,0) \nn
s_t &=& (0,\sin\vartheta_P \cos\varphi,
  \sin\vartheta_P \sin\varphi,\cos\vartheta_P)
\ena

The angular decay distribution in the transversity system can be obtained by
substituting the angle relations~(\ref{anglerel}) into the decay
distribution~(\ref{angdisthel}). One obtains
\bea\label{angdisttra}
W(\vartheta_P,\varphi) = A + B\ P_t\sin\vartheta_P \cos\varphi 
  + C\ P_t \sin\vartheta_P \sin\varphi + D\ P_t \cos\vartheta_P  
\ena

We conclude this section by taking a closer look at the two polar correlations
in helicity system I (\ref{angdisthel}) and the transversity system
(\ref{angdisttra}) where we include also the NLO QCD corrections as listed
e.g.\ in Ref.~\cite{Groote:2017pvc}. In helicity system I, one has
\be
\label{poldistr3}
W(\theta_P) \sim 1 + (1-O(\Delta))\, P_t \cos\theta_P 
\en
where $\Delta= (\delta^{(A)}-\delta^{(B)})/(\delta^{(A)}+\delta^{(B)})=0.00178$
quantifies the NLO corrections to the LO result $\Delta=0$. The values for
$\delta^{(A)}=A^{(1)}/A^{(0)}=-0.0846955$ and
$\delta^{(B)}=B^{(1)}/A^{(0)}=-0.0863048$ have been taken from
Ref.~\cite{Groote:2017pvc}. The NLO corrections to the LO distribution
$W(\theta_P) \sim 1 +\, P_t \cos\theta_P$ in Eq.~(\ref{poldistr3}) can be
seen to be very small even if one includes the non-SM couplings $\real f_L$
and $\real g_R$. 

In the transversity system one has the polar distribution 
\be\label{transpolar}
W(\vartheta_P) = 1 +\frac{1}{F^{(1)}(\real f_L,\real g_R)}\,
  \frac{3\pi(1-x^2)}{4(1+2x^2)}\,P_t\imag g_R\cos\vartheta_P 
\en
where
\bea
F^{(1)}(\real f_L, \real g_R)&=& 1+\delta^{(A)}+2\real f_L
-\frac{6x}{1+2x^2}\real g_R \nn
&\approx&(1+\delta^{(A)})\left(1+2\real f_L-\frac{6x}{1+2x^2}\real g_R\right)
  \nn
&=&(1+\delta^{(A)})F^{(0)}(\real f_L,\real g_R)
\ena

The usefulness of the transversity frame polar distribution is hampered by the
appearance of the unknown quantities $\real f_L$ and $\real g_R$ in the
denominator of Eq.~(\ref{transpolar}). As is frequently done when analyzing
the impact of more than one non-SM parameters on a given decay distribution
one adopts a strategy to allow one non-SM coupling at a time. For example, one
can set $\real f_L=0$ and $ \real g_R=0$ and keep only the non-SM coupling
$\imag g_R$. In this case, $F^{(1)}(\real f_L,\real g_R)=1+\delta^{(A)}$. One
finds that the analyzing power of the distribution~(\ref{transpolar}) is quite
large in that $3\pi(1-x^2)(1+\delta^{(A)})/4(1+2x^2)=1.41$. Since the
analyzing powers of both the helicity and transversity polar distributions are
quite large, this two-fold set of measurements must be judged to be a very
promising tool to simultaneously determine $P_t$ and $\imag g_R$. 

\section{\label{sec4}Positivity bounds in the helicity system}
First observe that the structure of the differential angular decay distribution
in helicity system I leaves little room for the contributions of the structure
functions $C$ and $D$ if the differential rate is to remain positive definite.
The LO differential decay distribution is given by
\be
\label{dist1}
\frac{1}{\Gamma}\frac{d\Gamma}{d\cos\theta d\phi} =
\frac{1}{4\pi}\bigg(A^{(0)}+B^{(0)} P_t \cos\theta_P
  + C^{(0)} P_t \sin\theta_p \cos\phi
  + D^{(0)} P_t \sin\theta_p \sin\phi \bigg)
\en
where $A^{(0)}=B^{(0)}$ in helicity system I. The LO polar analyzing structure
in helicity system I is maximal with $W(\theta_P)\sim 1+ P_t\cos\theta_p$. It
is heuristically clear that for $P_t=1$ one can immediately conclude that the
structure functions $C$ and $D$ must vanish as, in fact, is the case for the
LO values of $C^{(0)}$ and $D^{(0)}$. At NLO QCD the equality of $A$ and $B$
is slightly off-set where one now has $W(\theta_P)\sim 1+ 0.9982\cos\theta_p$
(setting again $P_t=1$) allowing for small contributions of $C$ and $D$.

Technically this is done by expanding $\cos\theta_P$ and $\sin\theta_P$ around
$\theta_P=\pi$ up to second order in $\delta=\pi\pm\theta_P$. The vanishing of
the discriminant of the corresponding quadratic equation defines the boundary
of the allowed values of the coefficients of the quadratic equation.

In Ref.~\cite{Groote:2017pvc}, this technique was applied to the
distribution~(\ref{dist1}) to derive a positivity bound on the $T$-odd
coupling factor $\imag g_R$. Including contributions from $\real f_L$ and
$\real g_R$, one has
\be
\label{boundhelI}
|\imag g_R|\le \frac{4(1+2x^2}{3\pi(1-x^2)}(1+\delta^{(A)})
  \sqrt{2\Delta(1-\Delta)}\sqrt{ F^{(0)}(\real f_L,\real g_R)}
\en
where we have used
\be
1+\delta^{(B)}+2\real f_L-\frac{6x}{1+2x^2}\real g_R\approx (1+\delta^{(B)})
  \left(1+2\real f_L-\frac{6x}{1+2x^2}\real g_R\right)
\en
For $\real f_L=\real g_R=0$ one has $F^{(0)}(\real f_L,\real g_R)=1$ and one
recovers the bound given in Ref.~\cite{Groote:2017pvc}. As noted above, at LO
one has $\Delta=0$ such that $\imag g_R=0$ at LO regardless of what values
$\real f_L$ and $\real g_R$ take.

Setting $\sin\phi=0$ in Eq.~(\ref{dist1}) one can derive a similar bound on
the $T$-even structure function $C$. One has
\be\label{boundC}
\Big|\Big(\frac{3\pi(1-x^2)}{4(1+2x^2}\real g_R - \delta^{(C)}\Big)
\Big|\le (1+\delta^{(A)})\sqrt{2\Delta(1-\Delta)}
\sqrt{ F^{(0)}(\real f_L,\real g_R)} 
\en
The NLO contribution $\delta^{(C)}$ to the T-even structure function $C$
appearing in Eq.~(\ref{boundC}) was calculated before in
Ref.~\cite{Groote:2006kq}. One has
\bea
\delta^{(C)}=\frac{C^{(1)}}{A^{(0)}}&=&-C_F\frac{\alpha_s}{4\pi}\frac34\pi
\frac{1}{(1-x^2)^2(1+2x^2)}\Big\{4x(4+3x^2-3x^4)(\Li_2(-x)-\Li_2(-1))\nn
&& -2(1-x^2)(8-7x+8x^2-5x^3)\ln(1+x)-\frac{(1-x^2)^3}{x}\ln(1-x^2)\nn[7pt]
&&+2x(1-x)^2(1-x-2x^2)\Big)\Big\}\ =\ -0.0024
\ena
As already demonstrated in Ref.~\cite{Groote:2006kq}, the NLO SM value for
$\delta^{(C)}=-0.0024$ easily satisfies the SM positivity bound given by
\be
|\delta^{(C)}|\le \Big(1+\delta^{(A)}\Big)\sqrt{2\Delta(1-\Delta)}=0.0542
\en

More straightforward bounds can be obtained from the various polar
distributions
\be
W(\theta_P)=(1+\kappa_i P_t \cos\theta_P)
\en
in form of the constraint $|\kappa_i| \le 1$ valid for $P_t=1$.
For example, for helicity system I one finds
\be
\kappa_{\rm I}=\Delta \frac{1}{F^{(1)}(\real f_L,\real g_R)}
\en
The corresponding bound is much weaker than the bound in
Eq.~(\ref{boundhelI}). For helicity system II one obtains 
($\delta_{\rm II}^{(B)}=B_{\rm II}^{(1)}/B_{\rm II}^{(0)}$)
\be
\kappa_{\rm II} =\frac{(1-2x^2)}{(1+2x^2)F^{(1)}(\real f_L,\real g_R)}
\left(1+\delta_{\rm II}^{(B)}+2\real f_L +\frac{2x}{1-2x^2}\real g_R\right)
\en
where, in system II, $\delta_{\rm II}^{(B)}= -0.12$
\cite{Fischer:1998gsa,Fischer:2001gp}. We do not explicitely list the
asymmetry parameter for helicity system III since the corresponding bound
is not very illuminating. Finally, the asymmetry parameter in the transversity
system reads
\be
\kappa_T=\frac{3\pi(1-x^2)}{4(1+2x^2)F^{(1)}(\real f_L,\real g_R)}
\en
Again, the bound resulting from $|\kappa_T| \le 1$ is much weaker than the
bound in Eq.~(\ref{boundhelI}).

Common to all the bounds discussed in this section is the necessity to prevent 
the denominator factor $F^{(1)}(\real f_L,\real g_R)$ from vanishing. This
gives a nontrivial restriction on the parameter space $(\real f_L,\real g_R)$
which would, for example, further restrict the bounds on $\real f_L$ and
$\real g_R$ derived from the weak radiative $B$ decays which read
$-0.13<\real f_L< 0.03$ and $-0.15< \real g_R< 0.57$~\cite{Grzadkowski:2008mf}.

\section{\label{sec5}Calculation of the imaginary contribution \\
  $\imag g_R$ from electroweak corrections}    
\begin{figure}\begin{center}
\epsfig{file=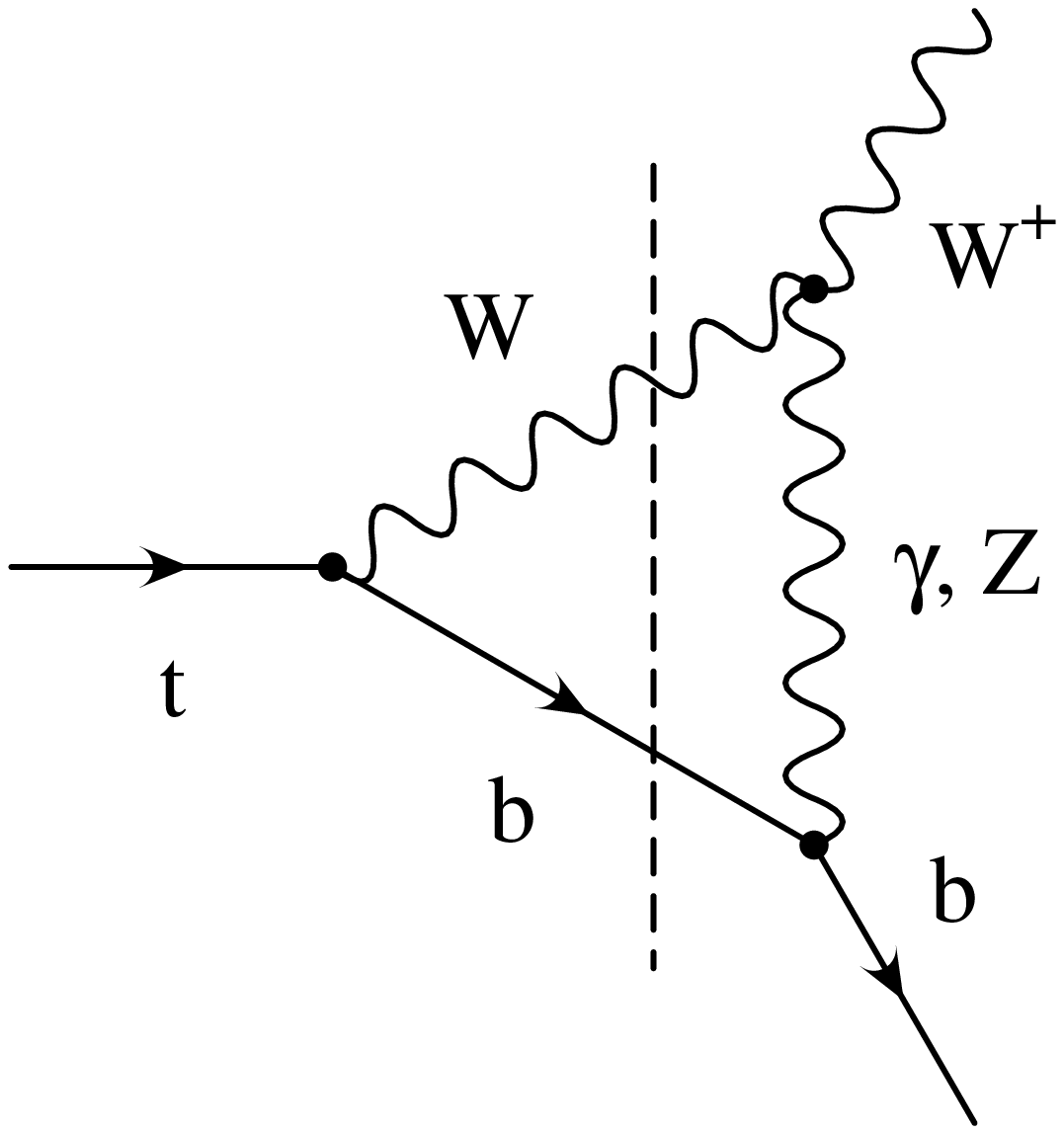, scale=0.35} \qquad
\epsfig{file=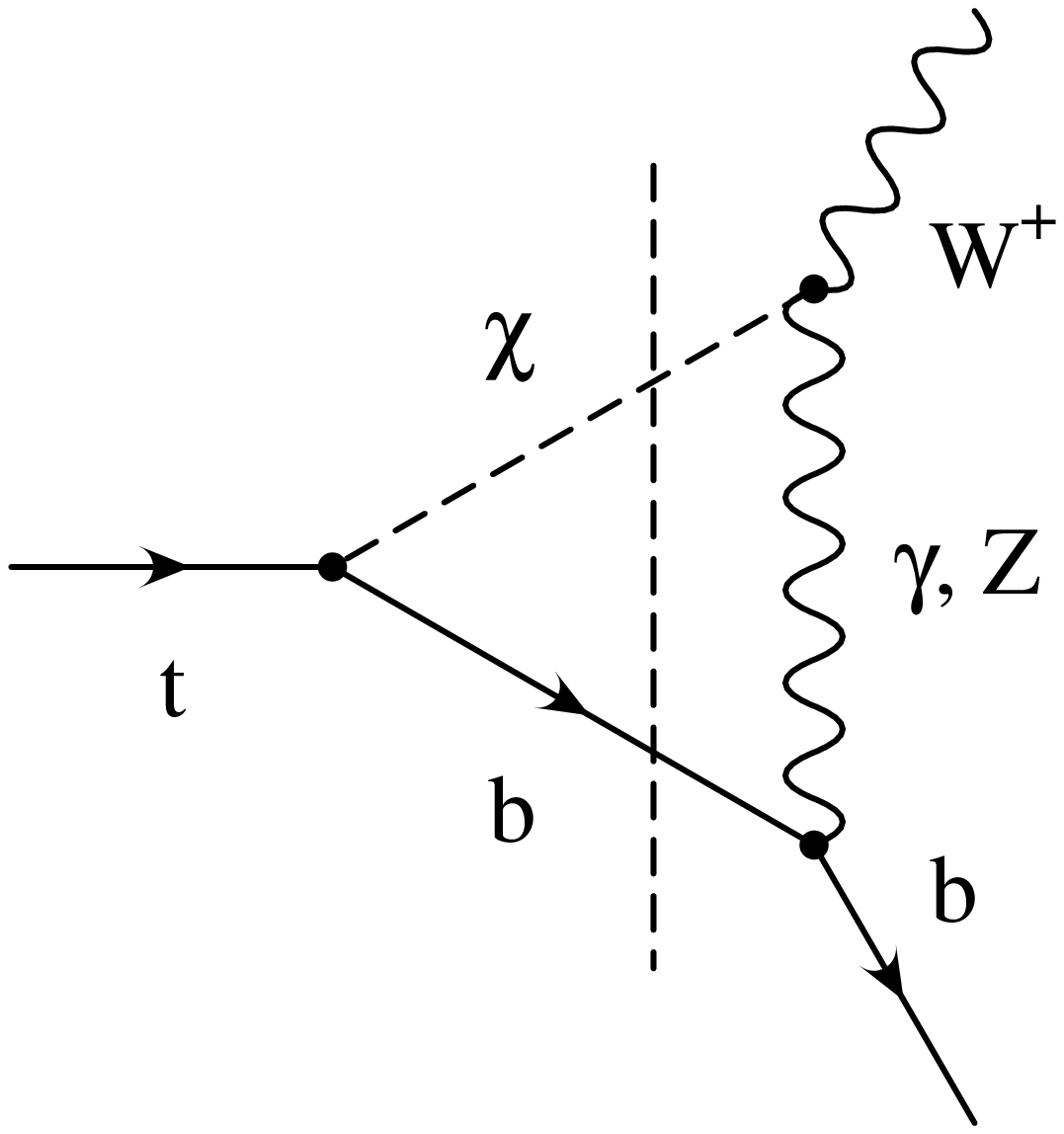, scale=0.35} 
\end{center}
\caption{\label{absorp}Absorptive parts of the four Feynman diagrams that
  contribute to $T$-odd correlations in polarized top quark decays}
\end{figure}

There are altogether 18 Feynman vertex diagrams that contribute to the decay
$t \to b+W^+$ at NLO of the electroweak interactions. Of these 18 diagrams,
seven diagrams admit absorptive parts. Three of these seven absorptive
diagrams give vanishing contributions for $m_b=0$. One finally remains with
four absorptive contributions which are depicted in Fig.~\ref{absorp}. In the
terminology of Ref.~\cite{GonzalezSprinberg:2011kx} the four diagrams are
labeled by $(A,B,C)=(b,W,\gamma(Z))$ and $(b,\chi,\gamma(Z))$. Note that the
contribution of the right diagram in Fig.~\ref{absorp} involving the Goldstone
boson $\chi$ is needed to render the on-shell gauge boson $W^+$ in the left
diagram to be four-transverse.

We have done a careful analysis of the absorptive parts of the diagrams in
Fig.~\ref{absorp} and their contributions to the two invariant amplitudes
$f_L$ and $g_R$. Note that there are no contributions to $f_R$ and $g_L$ in
the limit $m_b=0$. We have found $\imag f_L$ to be IR-divergent which is of
no concern since $\imag f_L$ does not contribute to physical observables at
NLO. The reason is that $\imag f_L$ multiplies the same covariance structure
as the Born term. In the following we concentrate on the evaluation of
$\imag g_R$. The imaginary part can be extracted from the logarithms
appearing in the loop calculation. As to be expected, $\imag g_R$ is infrared
and ultraviolet finite. The result is given by
\bea\label{im1}
\imag g_R(\gamma+Z) &=&\frac{\alpha}{4\pi}\,\,\bigg[Q_b\, \,x(2-x^2) 
  -\frac{(1+2Q_b\sin^2\theta_W)}{\sin^2\theta_W}\,\frac{1}{2x(1-x^2)^4}
  \Big\{ \nn
  && (1-x^2)^2\Big(x^2(1-x^2)^2(2-x^2)-2(1-3x^2-x^4)x_Z^2\Big) \nn
  && +\Big((1-x^2)^2(1-5x^2)x_Z^2+2(1-3x^2-x^4)x^4_Z\Big)\,\ell_Z
  \Big\}\bigg]\,\pi
\ena
where the numerically dominant logarithmic factor reads
\bea
\ell_Z&=& \ln \left(\frac{(x_Z^2+(1-x^2)^2)^2}
  {(x_Z^2-x^2(1-x^2))(x_Z^2+ (1-x^2)(1-2x^2))}\right)
\ena
The scaled masses of the $Z$ and $W$ boson are denoted by $x_Z=m_Z/m_t$ and
$x=m_W/m_t$, as before. The first term in Eq.~(\ref{im1}) proportional to
$Q_b=-1/3$ is due to $\gamma$ exchange while the remaining contribution is due
to $Z$ exchange. The analytical form of the $\gamma$-exchange contribution
agrees with the corresponding result in Ref.~\cite{GonzalezSprinberg:2011kx}
whereas the closed-form expression for the $Z$-exchange contribution in
Eq.~(\ref{im1}) is new.

Numerically one finds ($\alpha=1/128$, $\sin\theta_W=0.23126$,
$m_t=173.21\GeV$, $m_Z=91.1876$, $m_W=80.385$~\cite{Patrignani:2016xqp})
\bea
{\rm this\, \, calculation:} \qquad  \imag g_R (\gamma)&=&-0.539 \times 10^{-3}
   \qquad \imag g_R (Z)=-1.636 \times 10^{-3} \nn
\cite{GonzalezSprinberg:2011kx}:\qquad \qquad
   \imag g_R (\gamma)&=&-0.509 \times 10^{-3} \qquad
   \imag g_R (Z)=-0.726 \times 10^{-3}\nn
\cite{Arhrib:2016vts}:\qquad \qquad
    \imag g_R (\gamma)&=&-0.503 \times 10^{-3} \qquad
    \imag g_R (Z)=-1.601 \times 10^{-3} \nn
\ena
Up to small numerical differences we agree with
Refs.~\cite{GonzalezSprinberg:2011kx,Arhrib:2016vts} on the $\gamma$-exchange
contribution and with Ref.~\cite{Arhrib:2016vts} on the $Z$-exchange
contribution after taking into account that we are using a running
$\alpha(m_Z^2)=1/128$. The present calculation on the $Z$-exchange
contribution settles the factor 2 discrepancy between the results of
Ref.~\cite{GonzalezSprinberg:2011kx} and Ref.~\cite{Arhrib:2016vts} in favor
of the result of Ref.~\cite{Arhrib:2016vts}. The remaining small numerical
differences are very likely to result from inaccurate numerical integrations
in Refs.~\cite{GonzalezSprinberg:2011kx,Arhrib:2016vts}. Our combined result,
finally, is
\be
\imag\,g_R(\gamma+Z)=-2.175 \times 10^{-3}
\en
The result on $\imag\,g_R(\gamma+Z)$ is quite small. The result easily fits
into the experimental bound by ATLAS~\cite{Aaboud:2017aqp}
\be
\imag\,g_R \in [-0.18,\,0.06\,]
\en
and the theoretical positivity bound
\be
\imag\,g_R \in [-0.0420,\,0.0420\,]
\en
derived in Ref.~\cite{Groote:2017pvc}.

\section{\label{sec6}Summary and conclusion}
We have identified the $T$-odd structure functions that appear in the
description of polarized top quark decays and have written down the angular
factors that multiply them in the angular decay distribution. There are two
variants of angular decay distributions that have been used in the literature
to describe polarized top quark decays. These are the sequential decay
$t(\uparrow) \to X_b +W^+ (\to \ell^+ +\nu_\ell)$ and the quasi-three-body
decay $t(\uparrow) \to X_b+ \ell^+ + \nu_\ell$. The number of structure
functions needed to describe the quasi-three-body decay is smaller than the
number needed to describe the sequential decay. In this sense, the analysis of
the quasi-three-body decay $t(\uparrow) \to X_b+ \ell^+ + \nu_\ell$ constitutes
a more inclusive measurement than the analysis of the sequential decay
$t(\uparrow) \to X_b +W^+ (\to \ell^+ +\nu_\ell)$.

A convenient measure of the size of the $T$-odd contributions can be written
down in terms of the contribution of the imaginary part of the right-chiral
coupling $g_R$ appearing in the expansion of the general matrix element
$\langle b|J_{\rm eff}|t\rangle$. Contributions to $\imag g_R$ can either arise
from $CP$-violating interactions for which there is no SM source or from
$CP$-conserving final state interactions. In fact, within the  SM there exist
four NLO electroweak one-loop contributions which admit absorptive cuts. We
have provided analytical and numerical results for these absorptive cuts which
we present in terms of their contributions to $\imag g_R$. The size of these
absorptive contributions are rather small and easily fit into the existing
experimental~\cite{Aaboud:2017aqp,Aaboud:2017yqf} and
theoretical~\cite{Groote:2017pvc} bounds on $\imag g_R$.

We have elaborated on a possible simultaneous measurement of the polarization
of the top quark and $\imag g_R$ using a set of two independent polar decay
distributions involving the helicity and transversity systems in the
quasi-three-body decay. We have also commented on the bounds on the non-SM
coupling factors that result from the positivity of the differential angular
rate. To our knowledge these bounds have not been considered so far in global
analysis' of the allowed values of the non-SM coupling parameters
($\real f_L$, $\real g_R$, $\imag g_R$). In our analysis we have used the
$x_\ell$-integrated forms of the structure functions. It would be worthwhile
to similarly analyze the decay distributions and bounds using the unintegrated
forms of the structure functions. 

We mention that when going from top quark decays to antitop quark decays one
can distinguish the two sources of $CP$-violating phases. One has a phase
change $e^{i\chi} \to e^{-i\chi}$ for $CP$-violating phases and no phase
change $e^{i\chi} \to e^{i\chi}$ for $CP$-conserving final state interactions
where we assume that the final state interactions are $CP$-conserving (see
e.g.\ Refs.~\cite{Bernreuther:1992be,Bernreuther:2008us}).

\subsection*{Acknowledgments}
We would like to thank W.~Bernreuther, R.~Martinez, J.~Mueller and J.~Vidal
for e-mail exchanges on the subject of this paper. This work was supported by
the Estonian Science Foundation under Grant No.~IUT2-27. S.G.\ acknowledges
the hospitality of the theory group THEP at the Institute of Physics at the
University of Mainz and the support of the Cluster of Excellence PRISMA at the
University of Mainz. 


\end{document}